\documentclass[11pt]{article}
\DeclareMathAlphabet{\scr}{U}{rsfs}{m}{n}

\pdfoutput=1

\usepackage{latexsym}
\usepackage{epsfig}
\usepackage[mathscr]{eucal}
\usepackage{amsfonts}
\usepackage{amscd}
\usepackage{cite}
\usepackage{array}
\usepackage{bbold}
\usepackage{amssymb}
\usepackage{colordvi}
\usepackage[centertags]{amsmath}
\usepackage{enumerate}
\usepackage{graphicx}
\usepackage{booktabs}
\usepackage{theorem}
\usepackage[footnotesize]{caption}
\usepackage{soul}
\usepackage{url}
\usepackage[hidelinks]{hyperref}
\usepackage{mcite}
\usepackage{slashed}
\usepackage[dvipsnames]{xcolor}
\usepackage{bbm}
\usepackage[utf8]{inputenc}

\allowdisplaybreaks{}

\graphicspath{{analysis/}}
\newcommand{\newc}{\newcommand}
\newc{\bea}{\begin{eqnarray}}
\newc{\eea}{\end{eqnarray}}
\newc{\ol}{\overline}
\newc{\wt}{\widetilde}
\newc{\bs}{\boldsymbol}
\newc{\m}{\mathcal}
\newc{\la}{\langle}
\newc{\ra}{\rangle}

\newcommand{\beq}{\begin{eqnarray}}
\newcommand{\eeq}{\end{eqnarray}}
\newcommand{\bpmatrix}{\begin{pmatrix}}
\newcommand{\epmatrix}{\end{pmatrix}}

\renewcommand{\ol}{\text{1l}}

\renewcommand{\eqref}[1]{Eq.~(\ref{#1})}

\newcommand{\bc}{\begin{center}}
\newcommand{\ec}{\end{center}}

\newenvironment{Eqnarray}%
     {\arraycolsep 0.14em\begin{eqnarray}}{\end{eqnarray}}
\newcommand{\ba}{\begin{Eqnarray}}
\newcommand{\ea}{\end{Eqnarray}}
\newcommand{\be}{\begin{equation}}
\newcommand{\ee}{\end{equation}}

\begin{document}

\title{
\vspace*{-3.7cm}
\phantom{h} \hfill\mbox{\small }
\\[1cm]
\textbf{Vacuum structure of the $\mathbb{Z}_2$ symmetric Georgi-Machacek model \\[4mm]}}

\date{}
\author{
Duarte Azevedo$^{1,\,}$,
Pedro Ferreira$^{1, 2\,}$\footnote{E-mail:
\texttt{pmmferreira@fc.ul.pt}} ,
Heather E.~Logan$^{3\,}$\footnote{E-mail:
\texttt{logan@physics.carleton.ca}} ,
Rui Santos$^{1,2\,}$\footnote{E-mail:
  \texttt{rasantos@fc.ul.pt}} ,
\\[5mm]
{\small\it
$^1$Centro de F\'{\i}sica Te\'{o}rica e Computacional,
    Faculdade de Ci\^{e}ncias,} \\
{\small \it    Universidade de Lisboa, Campo Grande, Edif\'{\i}cio C8
  1749-016 Lisboa, Portugal} \\[3mm]
{\small\it
$^2$ISEL -
 Instituto Superior de Engenharia de Lisboa,} \\
{\small \it   Instituto Polit\'ecnico de Lisboa
 1959-007 Lisboa, Portugal} \\[3mm]
{\small\it
$^3$Ottawa-Carleton Institute for Physics, Carleton University,}\\
{\small\it 1125 Colonel By Drive, Ottawa, Ontario K1S 5B6, Canada}\\[3mm]
}
\maketitle

\begin{abstract}
We discuss the vacuum structure of a version of the Georgi-Machecek model with an exact  $\mathbb{Z}_2$ symmetry acting on the triplet fields. Besides
the usual custodial-symmetric model, with $\rho=1$ at tree-level, a model with a dark matter candidate is also viable. The other phases of the model lead
to electric charge breaking, a wrong pattern of electroweak symmetry breaking or to $\rho \neq 1$ at tree-level. We derive conditions to have
an absolute minimum in each of the two viable phases, the custodial and the dark matter phases.
\end{abstract}
\thispagestyle{empty}
\vfill
\newpage
\setcounter{page}{1}


\maketitle

\section{Introduction}

The Georgi-Machacek~\cite{Georgi:1985nv} (GM) model was proposed in 1985 in an attempt to understand if  patterns of electroweak symmetry breaking different than the one
proposed in the Standard Model (SM) were still viable. Besides the SM complex doublet, the model has one real triplet and one complex triplet.
Clearly the model leads to different phenomenology at colliders but it is also able to reproduce all the available experimental results. It is of particular importance,
because it is not trivial, that the SM ratio of charged to neutral currents can be obtained at tree-level. This ratio is translated
into the relation $\rho= m_W^2/(m_Z^2 \cos^2 \theta_W) = 1$ at tree-level ($m_{W (Z)}$ is the $W(Z)$ boson mass) and is measured with great precision
allowing only for small deviations from higher order corrections.
So far experiments have neither confirmed nor excluded the spontaneous breaking pattern predicted by the SM. Hence, all possibilities which agree
with the observed phenomenology should be considered.

In the same year a Higgs potential was written explicitly for the GM model by Chanowitz and Golden~\cite{Chanowitz:1985ug}. The potential has only terms with mass dimension
of order 2 and order 4 and is invariant under a global $SU(2)_L \times SU(2)_R$ symmetry. For this particular version of the model the vacuum expectation values of the
neutral components of the triplet can be arranged such that a custodial $SU(2)$ symmetry survives after the spontaneous breaking of the global $SU(2)_L \times SU(2)_R$ symmetry.
This way, the equality of the vacuum expectation values of the neutral fields which leads to $\rho=1$ at tree-level is naturally preserved in the Higgs sector. 

Several versions of GM model and its phenomenology have been discussed over the years~\cite{Gunion:1989ci, Gunion:1990dt, Cheung:1994rp, Aoki:2007ah, Logan:2010en,
Chiang:2012cn, Englert:2013wga, Chiang:2013rua, Kanemura:2014bqa, Hartling:2014zca, Chiang:2014bia, Degrande:2015xnm, Chang:2017niy, Keeshan:2018ypw, Ghosh:2019qie}  and several searches were proposed and performed at colliders and in particular at the Large Hadron Collider (LHC).
The latest constraints on the GM model can be found in~\cite{Chiang:2018cgb, Ismail:2020zoz}.
There is also  a public code available, GMCALC~\cite{Hartling:2014xma}, which allows to calculate the particle spectrum of the model together
with the couplings and decay widths. Checks of perturbative unitarity and boundedness from below are included in the code.  Starting from a parameter point that constitutes a viable custodial-symmetric minimum of the scalar potential, the code also checks for deeper alternative minima by performing a numerical scan following the method proposed in Ref.~\cite{Hartling:2014zca}.  A full study of the possible vacua of the model has not been performed.

In this work we discuss the vacuum structure of the $\mathbb{Z}_2$ symmetric model as proposed in ref.~\cite{Chanowitz:1985ug} which is
a simple yet viable version of the GM potential. The model with this potential and no further field additions has a
Dark Matter (DM) phase when
spontaneous symmetry breaking leaves the $\mathbb{Z}_2$ unbroken, that is, when the triplets have no vacuum expectation
values (VEVs). The model studied in the literature
is the one where custodial $SU(2)$ is preserved and $\rho=1$ at tree-level. The model has however other types of minima that do not preserve
the custodial symmetry that range from electric charge breaking minima to models with just the wrong pattern of spontaneous symmetry breaking (SSB). Besides, these non-custodial vacua always give rise to extra Goldstone bosons.  We will compare the depths of the potential calculated at the different vacua at tree-level to understand
if either the DM or the Custodial vacuum are naturally stable and derive the conditions for absolute stability in case they are not.  The expressions found can be used to ensure that the desired vacuum is the deepest one at tree-level.

In this first attempt to understand the vacuum structure of the model we make the analysis more tractable by considering only the real component of the (in general complex) VEVs. In that sense the conditions we obtain for the potential to be in an absolute minimum are necessary conditions.

The paper is organized as follows. In section~\ref{sec:model} we briefly present the potential and the vacuum structure of the model and
in section~\ref{BilForm} we present the bilinear formalism. Models that retain the custodial symmetry after spontaneous symmetry breaking (SSB)
are discussed in section~\ref{sec:cust} while the ones where said symmetry is broken are discussed in~\ref{sec:noncust}.  In section~\ref{sec:viable}
we discuss the stability of the two viable models. We summarise our findings in section~\ref{sec:Conc}.

\section{Scalar potential and vacuum structure}
\label{sec:model}
The Georgi-Machacek model~\cite{Georgi:1985nv} consists of the usual complex doublet $(\phi^+,\phi^0)$ with
hypercharge, $Y$, equal to 1, a real triplet $(\xi^+,\xi^0,\xi^-)$ with $Y = 0$, and  a complex triplet $(\chi^{++},\chi^+,\chi^0)$ with $Y=2$. The fields can be written in the following matrix form
\begin{eqnarray}
	\Phi &=& \left( \begin{array}{cc}
	\phi^{0*} &\phi^+  \\
	-\phi^{+*} & \phi^0  \end{array} \right), \\
	\Xi &=&
	\left(
	\begin{array}{ccc}
	\chi^{0*} & \xi^+ & \chi^{++} \\
	 -\chi^{+*} & \xi^{0} & \chi^+ \\
	 \chi^{++*} & -\xi^{+*} & \chi^0
	\end{array}
	\right).
	\label{eq:PX}
\end{eqnarray}
which transform under the global SU(2)$_L \times$SU(2)$_R$ symmetry according to
\begin{equation}
	\exp(i T^a \theta_L^a) (\Phi \ {\rm or} \ \Xi) \exp(-i T^b \theta_R^b).
\end{equation}

The most general scalar potential invariant under the global SU(2)$_L \times$SU(2)$_R$ symmetry is
\begin{eqnarray}
	V(\Phi,\Xi) &= & \frac{\mu_2^2}{2}  {\rm Tr}(\Phi^\dagger \Phi)
	+  \frac{\mu_3^2}{2}  {\rm Tr}(\Xi^\dagger \Xi)
	+ \lambda_1 [{\rm Tr}(\Phi^\dagger \Phi)]^2
	+ \lambda_2 {\rm Tr}(\Phi^\dagger \Phi) \text{Tr}(\Xi^\dagger \Xi)   \nonumber \\
          & & + \lambda_3 {\rm Tr}(\Xi^\dagger \Xi \Xi^\dagger \Xi)
          + \lambda_4 [{\rm Tr}(\Xi^\dagger \Xi)]^2
           - \lambda_5 {\rm Tr}( \Phi^\dagger \tau^a \Phi \tau^b) \text{Tr}( \Xi^\dagger t^a \Xi t^b)
           \label{eq:potential}
\end{eqnarray}
where we have further imposed a $\mathbb{Z}_2$ symmetry, $\Xi \to -\Xi$, which eliminates the cubic terms ${\rm Tr}(\Phi^{\dagger} \tau^a \Phi \tau^b) (U \Xi U^{\dagger})_{ab}$
and ${\rm Tr}(\Xi^{\dagger} t^a \Xi t^b) (U \Xi U^{\dagger})_{ab}$ (for the definition of $U$ see Ref.~\cite{Aoki:2007ah}). This is the potential originally proposed by Chanowitz and Golden~\cite{Chanowitz:1985ug}
which not only preserves $\rho = 1$ at tree-level but also offers a new possibility that the triplets have the dominant contribution to the W boson mass.
The potential must be bounded from below,
and the necessary and sufficient conditions that the quartic couplings $\lambda_i$ must obey for that
to happen have been established~\cite{Hartling:2014zca,Hartling:2014xma}:
\begin{eqnarray}
	\lambda_1 &>& 0, \nonumber \\
	\lambda_4 &>& \left\{ \begin{array}{l l}
		- \frac{1}{3} \lambda_3 & {\rm for} \ \lambda_3 \geq 0, \\
		- \lambda_3 & {\rm for} \ \lambda_3 < 0, \end{array} \right. \nonumber \\
	\lambda_2 &>& \left\{ \begin{array}{l l}
		\frac{1}{2} \lambda_5 - 2 \sqrt{\lambda_1 \left( \frac{1}{3} \lambda_3 + \lambda_4 \right)} &
			{\rm for} \ \lambda_5 \geq 0 \ {\rm and} \ \lambda_3 \geq 0, \\
		\omega_+(\zeta) \lambda_5 - 2 \sqrt{\lambda_1 ( \zeta \lambda_3 + \lambda_4)} &
			{\rm for} \ \lambda_5 \geq 0 \ {\rm and} \ \lambda_3 < 0, \\
		\omega_-(\zeta) \lambda_5 - 2 \sqrt{\lambda_1 (\zeta \lambda_3 + \lambda_4)} &
			{\rm for} \ \lambda_5 < 0 \, ,
			\end{array} \right.
	\label{eq:bfb}
\end{eqnarray}
with
\begin{equation}
	\omega_{\pm}(\zeta) = \frac{1}{6}(1 - B) \pm \frac{\sqrt{2}}{3} \left[ (1 - B) \left(\frac{1}{2} + B\right)\right]^{1/2},
\end{equation}
and
\begin{equation}
	B \equiv \sqrt{\frac{3}{2}\left(\zeta - \frac{1}{3}\right)} \in [0,1]; \qquad  \zeta \in \left[ \frac{1}{3}, 1 \right] .
\end{equation}

Due to the hermiticity of the potential, and its field structure, all of the couplings in~\eqref{eq:potential}
are necessarily real, and therefore the CP symmetry cannot be explicitly broken.
As we will see later any attempt of spontaneous CP breaking would lead to the breaking of the custodial symmetry.

As we are interested in the vacuum structure of the GM model we start by writing the vacuum expectation values (VEVs) of the fields in the following generic form
\begin{equation}
	\Phi = \frac{1}{\sqrt{2}}\left( \begin{array}{cc}
	v_1 - i v_2 & v_3 + i v_4  \\
	-v_3 + i v_4 & v_1 + i v_2  \end{array} \right), \qquad
	\Xi = \frac{1}{\sqrt{2}}
	\left(
	\begin{array}{ccc}
	v_8 - i v_9 & v_6 + i v_7 & v_{12} + i v_{13} \\
	 -v_{10} + i v_{11} & \sqrt{2} v_5 & v_{10} + i v_{11} \\
	 v_{12} - i v_{13} & -v_6 + i v_7 & v_8 + i v_9
	\end{array}
	\right).
	\label{eq:vevs}
\end{equation}
There are 6 symmetry transformations in SU(2)$_L \times$SU(2)$_R$ that can be used to eliminate the redundant VEVs.  We first use the three broken symmetry transformations, $\theta_R^a = -\theta_L^a$ (these are the ones orthogonal to the custodial SU(2)), to rotate away three of the four components of the doublet, leaving only $v_1$.  We can then use the three custodial symmetry transformations, $\theta_R^a = \theta_L^a$, to rotate away three of the VEVs in the triplet.  There are several possibilities and we choose to eliminate both the real and imaginary components of $\chi^{++}$ and the imaginary component of $\xi^+$, which leads to the following form of the vacuum configuration
\begin{equation}
	\Phi =  \frac{1}{\sqrt{2}} \left( \begin{array}{cc}
	v_1 & 0  \\
	0 & v_1  \end{array} \right), \qquad \qquad
	\Xi =\frac{1}{\sqrt{2}}
	\left(
	\begin{array}{ccc}
	v_8 - i v_9 & v_6 & 0 \\
	 -v_{10} + i v_{11} & \sqrt{2} v_5 & v_{10} + i v_{11} \\
	 0 & -v_6 & v_8 + i v_9
	\end{array}
	\right).
	\label{eq:vevs2}
\end{equation}
We are therefore left with 7 non-redundant VEVs. The scalar potential in Eq.~(\ref{eq:potential}) contains 4 distinct invariants under the SU(2)$_L \times$SU(2)$_R$ global symmetry.  These were identified in Ref.~\cite{Hartling:2014zca} and we write them as
\begin{eqnarray}
	{\rm Tr}(\Phi^{\dagger} \Phi) &=& r^2 \cos^2 \gamma, \\
	{\rm Tr}(\Xi^{\dagger} \Xi) &=& r^2 \sin^2 \gamma, \\
	{\rm Tr}(\Xi^{\dagger} \Xi \Xi^{\dagger} \Xi) &=& \zeta \cdot (r^2 \sin^2 \gamma)^2, \\
	{\rm Tr}(\Phi^\dagger \tau^a \Phi \tau^b) \text{Tr}(\Xi^\dagger t^a \Xi t^b) &=& \omega \cdot (r^2 \cos^2 \gamma)(r^2 \sin^2 \gamma)  \, .
\end{eqnarray}
The fact that the potential can be expressed in a four-dimensional subspace spanned by these four invariants implies that there are three extra symmetries besides SU(2)$_L \times$SU(2)$_R$ that have not yet been identified.

\section{Bilinear formalism and mass matrices for the GM model}
\label{BilForm}
In this section we will outline how a bilinear formalism may be applied to the study of the vaccum
structure of the GM model. In short, such a formalism expresses the potential, and its minimisation
conditions, in terms of quadratic functions of its fields and/or VEVs, which has been proven
to be an effective strategy to compare the values of scalar potentials at different stationary
points, and ascertain their relative depths. In the context of the two-Higgs doublet model (2HDM),
a first version of the bilinear formalism was first applied
in~\cite{Velhinho:1994np,Ferreira:2004yd,Barroso:2005sm}, and then refined
in~\cite{Nishi:2006tg,Maniatis:2006fs,Ivanov:2006yq,
  Barroso:2007rr,Nishi:2007nh,Maniatis:2007vn,Ivanov:2007de,Maniatis:2007de,Nishi:2007dv,
  Maniatis:2009vp,Ferreira:2010hy}, allowing for the study of the 2HDM's vacuum structure and possible
 global symmetries. The bilinear formalism was also applied to models with
different scalar content, to wit the 3HDM
\cite{Ivanov:2010ww,Ivanov:2014doa}, the complex singlet-doublet model (CXSM)
\cite{Ferreira:2016tcu}, the 2HDM with a real or complex singlet
(N2HDM)~\cite{Ferreira:2019iqb,Engeln:2020fld}, and the Higgs triplet model (HTM)~\cite{Ferreira:2019hfk}.
For most models, the bilinears are simply constructed as the four gauge-invariant quantities allowed with the fields present; but in the case of the HTM, the bilinears were simply the most convenient
quadratic products of VEVs to express the value of the potential at generic stationary points as a
quadratic polynomial. As we will shortly see, the same holds for the GM model --- the presence of a
triplet, as in~\cite{Ferreira:2019hfk}, yields such a structure of the
potential that we have not found it possible to express it as a product of quadratic gauge invariant field
products.

First however, in order to identify whether a given stationary point is a minimum, we will
need to study the second derivatives of the potential.
In order to obtain the mass matrices for all minima simultaneously we will express the doublet $\Phi$
and the triplet $\Xi$ of~\eqref{eq:PX} in terms of the real components of the fields $\varphi_i$,
according to
\begin{eqnarray}
	\Phi &=& \frac{1}{\sqrt{2}} \left( \begin{array}{cc}
	\varphi_1 - i \varphi_2 & \varphi_3 + i \varphi_4   \\
	-\varphi_3 + i \varphi_4 & \varphi_1 + i \varphi_2   \end{array} \right), \\
	\Xi &=&
	\frac{1}{\sqrt{2}}\left(
	\begin{array}{ccc}
	\varphi_8 - i \varphi_9  & \varphi_6 + i \varphi_7  & \varphi_{12} + i \varphi_{13}  \\
	 -\varphi_{10} + i \varphi_{11}  & \sqrt{2}\varphi_5  & \varphi_{10} + i \varphi_{11} \\
	 \varphi_{12} - i \varphi_{13} & -\varphi_6 + i \varphi_7 & \varphi_8 + i \varphi_9
	\end{array}
	\right).
\end{eqnarray}
With these conventions, the matrix elements of the $13\times
13$ mass matrix are
\begin{equation}
	\left[M^2\right]_{ij}\,=\,\frac{\partial^2 V}{\partial \varphi_i \partial \varphi_j}\,,
\end{equation}
with the potential given in~\eqref{eq:potential}. In order to make the analysis tractable
we will only consider the real part of the VEVs from~\eqref{eq:vevs2}, that is,
\begin{equation}
	\langle \varphi_1\rangle = v_1\;,\;\langle \varphi_5\rangle = v_5\;,\;\langle \varphi_6\rangle = v_6\;,\;
	\langle \varphi_8\rangle = v_8 \;,\; \langle \varphi_{10}\rangle=v_{10} \;,\;
	\label{eq:vevger}
\end{equation}
with $v_9$ and $v_{11}$ set equal to zero.  We leave a full analysis including these imaginary parts of the vevs to future work. This leads to a potential (evaluated at a stationary point for which these VEVs hold) equal to
\begin{equation}
\begin{aligned}
V&=v_1^2 \left[\frac{\mu_2^2}{2}+\lambda_1 v_1^2+\lambda_2 v_5^2+ \left(\lambda_2-\frac{\lambda_5}{4}\right)v_8^2+\lambda_2
v_6^2+\lambda_2 c_{10}^2-\frac{\lambda_5 v_5 v_8}{\sqrt{2}}-\frac{\lambda_5 v_{10}
	v_6}{2}\right]\\
&+v_5^2 \left[\frac{\mu_3^2}{2}+(\lambda_3+\lambda_4)v_5^2 +2 \lambda_4 v_8^2 + (2 \lambda_3+2 \lambda_4)v_6^2+(2 \lambda_3+2 \lambda_4)v_{10}^2 \right]\\
&+v_8^2 \left[\frac{\mu_3^2}{2}+
\left(\frac{\lambda_3}{2}+\lambda_4\right)v_8^2+ (\lambda_3+2 \lambda_4)v_6^2+(\lambda_3+2 \lambda_4)v_{10}^2 \right]\\
&+v_6^2 \left[\frac{\mu_3^2}{2}+ (\lambda_3+\lambda_4)v_6^2+2 \lambda_4 v_{10}^2\right]+v_{10}^2\left[\frac{\mu_3^2}{2}+ (\lambda_3+\lambda_4)v_{10}^2\right]-2 \sqrt{2} \lambda_3 v_{10} v_5 v_6
v_8.
\end{aligned}
\label{eq:potencial_real_vevs}
\end{equation}

Let us now define the vector $A$ and the symmetric matrix $B$ as
\begin{equation}
	A = \left(\begin{array}{c} \mu_2^2 \\ \mu_3^2 \\ 0 \\ 0 \\ 0 \\ 0
	\end{array}\right)\quad\quad ,\quad\quad
	B = \left(
	\begin{array}{cccccc}
	8 \lambda_1 & 4 \lambda_2 & -\lambda_5 &
	-\lambda_5 & -\lambda_5 & 0 \\
	4 \lambda_2 & 8 (\lambda_3+\lambda_4) & 0 & 0 & 0 & 0
	\\
	-\lambda_5 & 0 & -2 \lambda_3 & 0 & -2 \lambda_3 & 0
	\\
	-\lambda_5 & 0 & 0 & -4 \lambda_3 & 0 & -2 \lambda_3
	\\
	-\lambda_5 & 0 & -2 \lambda_3 & 0 & -4 \lambda_3 & 0
	\\
	0 & 0 & 0 & -2 \lambda_3 & 0 & 0 \\
	\end{array}
	\right) ,
	\label{eq:def}
\end{equation}
and the vector $X$, which depends on the VEVs as
\begin{equation}
	X=\left(
	\begin{array}{c}
	v_1^2/2 \\
	 \left(v_{10}^2+v_5^2+v_6^2+v_8^2\right)/2 \\
	\sqrt{2} v_5 v_8 \\
	v_8^2/2 \\
	v_{10} v_6 \\
	v_{10}^2+v_6^2 \\
	\end{array}
	\right).
	\label{eq:Xvector}
\end{equation}
With these definitions the value of the potential at each stationary point (SP) with VEVs $v_i$, is given by
\be
V_{SP}\;=\;A^T X_{SP}\,+\,\frac{1}{2}\, X^T_{SP}\,B\,X_{SP}\,.
\label{eq:VSP}
\ee

Thus we see that it is possible to express the value of the potential at a given
 stationary point as a quadratic polynomial on $X$, that is, an order 2 polynomial expressed
 in terms of bilinears, quadratic combinations of vevs~\footnote{Though it is not possible, due to the
 $\lambda_3$ and $\lambda_5$ terms in the potential, to express it as a quadratic polynomial of
 quadratic products of fields.}.
Further, since the potential is the sum of two homogeneous functions of order 2 and 4
($V_2$ and $V_4$, the quadratic and quartic terms in the vevs -- linear and quadratic in $X$), we have
\be
\mbox{At any stationary point:}\, \frac{\partial V}{\partial \varphi_i}\,=\,0
\,\Longrightarrow\,\sum_i \varphi_i\,\frac{\partial V}{\partial \varphi_i}\,=\,0\,\Longrightarrow
2\,V_2\,+\,4\,V_4\,=\,0\,.
\label{eq:homog}
\ee
The value of the potential at a given stationary point, $V_{SP}$, is simply
\be
V_{SP}\;=\;\frac{1}{2}\,\left(V_2\right)_{SP}\;=\;-\,\left(V_4\right)_{SP}\;=\;\frac{1}{2}\,A^T X_{SP}
\;=\;-\, X^T_{SP}\,B\,X_{SP}\,.
\label{eq:Vred}
\ee
Finally, we also need the vector $V^\prime$, defined as
\be
V^\prime\;=\; \frac{\partial V}{\partial X^T}\;=\; A\,+\,B\,X\, ,
\label{eq:Vl}
\ee
and since $V^\prime$ is the gradient of the potential along the vector $X$, the condition $X^T V^\prime = 0$ has to hold.
Let us now consider two distinct vacua with vectors $X_1$ and $X_2$, respectively. Let $V_1'$ and $V_2'$ be the gradient of the potential considered at those stationary points.
We can contract the vectors of the VEVs from one minimum with the gradient of the other to obtain
\begin{equation}
	X_1^T V_2'=X_1^T A+ X_1^T B X_2.
\end{equation}
Since $B$ is symmetric, we have
\begin{equation}
	X_1^T B X_2=X_2^T B X_1,
\end{equation}
which leads to
\begin{equation}
	X_2^T V_1' - X_2^T A = X_1^T V_2' - X_1^T A.
	\label{eq:prefinal}
\end{equation}
The second term in each side of the equation is the quadratic part of the potential computed at the corresponding stationary point. Therefore using~\eqref{eq:Vred} we can rewrite~\eqref{eq:prefinal} as
\begin{equation}
	V_2 - V_1 = \frac{1}{2}\left( X_2^TV_1'-X_1^TV_2' \right),
\end{equation}
which relates the relative height between the two minima with the VEV vectors and gradients of the potential.
We want to express the right hand side for each case as a function of physical parameters from one vacuum as much as possible. In this way, the stability of that vacuum is just a function of quantities computed at that stationary point. In the next section we will classify the different type of stationary points and write the corresponding $X$ and $V'$ vectors.
The minimum conditions that led to the different phases to be presented
are shown in appendix~\ref{app:vac}. The vacua classified as charge breaking are the ones where the photon becomes massive and the mass
matrix for the gauge bosons is also given in appendix~\ref{app:vac}.

\section{Models with custodial symmetry after SSB}
\label{sec:cust}
\subsection{The custodial vacuum}

The GM model was first proposed due to the very nice property that there is a solution of the minimization conditions
with no charged VEVs where the triplet neutral VEVs (those of the real fields $\varphi_5$ and $\varphi_8$)
are proportional. This condition preserves the custodial symmetry and maintains the SM structure of the gauge bosons mass matrix, that is, the
Weinberg angle is preserved at tree-level. As discussed in the introduction,  the equality of vacuum expectation values is naturally preserved
in the Higgs sector because the custodial $SU(2)$ survives the spontaneous breaking of the global
$SU(2)_L \times SU(2)_R$ symmetry.
In this scenario the VEVs are chosen as
\begin{equation}
\langle \varphi_1\rangle = v_1\;,\;\langle \varphi_5\rangle = v_5\;,\;\langle \varphi_6\rangle = 0\;,\;
\langle \varphi_8\rangle = \sqrt{2} v_5 \;,\; \langle \varphi_{10}\rangle=0 \;,\;
\label{eq:vevCUS}
\end{equation}
leading to the minimization conditions
\begin{equation}
\begin{aligned}
\mu_2^2&=-4 v_1^2 \lambda_1+ 3 v_5^2 (\lambda_5-2 \lambda_2)\, ,\\
\mu_3^2&=v_1^2 (\lambda_5-2 \lambda_2)-4 v_5^2
(\lambda_3+3 \lambda_4)\, .
\end{aligned}
\end{equation}
The mass spectrum is the following: there are two custodial singlet states that mix, $h$ and $H$, with masses given by
\begin{eqnarray}
	m_{h,\, H}^2 & = & 4 \lambda_1 v_1^2+4
	v_5^2 (\lambda_3+3 \lambda_4) \mp \nonumber \\
	&&  2 \sqrt{4 \lambda_1^2 v_1^4+v_1^2 v_5^2 \left[3 (\lambda_5-2 \lambda_2)^2-8 \lambda_1 (\lambda_3+3 \lambda_4)\right]+4 v_5^4 (\lambda_3+3 \lambda_4)^2} \, ,
\label{eq:mHh_C}
\end{eqnarray}
a custodial triplet, $H_3^0,H_3^\pm$, with masses given by
\begin{equation}
	m_3^2=\frac{1}{2} \lambda_5 \left(v_1^2+8 v_5^2\right) \, ,
\label{eq:m3_C}
\end{equation}
and a custodial 5-plet, $H_5^0, H_5^\pm, H_5^{\pm\pm}$, with masses
\begin{equation}
	m_5^2=\frac{3 \lambda_5 v_1^2}{2}+8 \lambda_3 v_5^2 \, .
\label{eq:m5_C}
\end{equation}
In order that the gauge bosons have the observed mass the VEVs have to obey
\begin{equation}
v_1^2 + 8 v_5^2 = v^2 = \frac{1}{\sqrt{2} G_F} \approx (246 \, {\rm GeV})^2 \,.
\end{equation}
There are also two mixing angles usually taken as input parameters and that parametrize the mixing between the doublet and the triplet components~\cite{Gunion:1989ci}.

Since the cubic terms are absent due to the $\mathbb{Z}_2$ symmetry, $\Xi \to -\Xi$, the mass scale of the scalar states
is set by the electroweak scale and the model does not possess a decoupling limit.
The upper bounds on the scalar masses from perturbative unitarity were first studied in Ref.~\cite{Aoki:2007ah} which found the following constraints on the masses: $m_3 \lesssim 400$~GeV, $m_5 \lesssim 700$~GeV.  Despite the absence of a decoupling limit, the model in this phase is still not excluded by data~\cite{Das:2018vkv}.

Finally, the vectors $X$ and $V^\prime$ can be written as
\begin{equation}
	X_{C}=\left(
	\begin{array}{c}
		v_1^2/2 \\
		3 v_5^2/2 \\
		2 v_5^2 \\
		v_5^2 \\
		0 \\
		0 \\
	\end{array}
	\right),
	\qquad V'_{C}=A+ B X_{C}=\left(
	\begin{array}{c}
	0 \\
	2\bar{m}^2 \\
	-\bar{m}^2 \\
	-\bar{m}^2 \\
	-\bar{m}^2 \\
	-2 \lambda_3 v_5^2 \\
	\end{array}
	\right) ,
\end{equation}%
with $\bar{m}^2= \frac{1}{2}\lambda_5 v^2_1 +4\lambda_3\,v_5^2$.
%

\subsection{The dark matter vacuum}
\label{sec:dm}

If only the doublet $\Phi$ develops a VEV, the Lagrangian will still be exactly $\mathbb{Z}_2$ symmetric after SSB.
This in turn means that all the particles from the triplet do not interact with
fermions nor do they have triple vertices with two gauge bosons. The lightest of these particles is thus stable and is therefore a good dark matter
candidate. Such a situation is indeed a possible solution of the minimization equations of the potential,
corresponding to VEVs $v_1\equiv v$ and all others zero, in \eqref{eq:vevger}. The only minimization condition is
\begin{equation}
	v = \sqrt{- \frac{\mu_2^2}{4 \lambda_1}} \, .
\label{eq:vevdm}
\end{equation}
In this scenario the custodial symmetry is still preserved and therefore the physical particles fall into the same custodial representations as in the custodial vacuum.  The two custodial singlet states no longer mix; the 125~GeV Higgs boson is identified with $h = \varphi_1$ with mass given by
\begin{equation}
	m_h^2 = 8 \lambda_1 v^2.
	\label{eq:mhdm}
\end{equation}
The $\mathbb{Z}_2$-odd states comprise the familiar custodial singlet, triplet, and fiveplet with masses given by
\begin{eqnarray}
	m_1^2 &=& (\mu_3^2 + 2 \lambda_2 v^2) - \lambda_5 v^2, \nonumber \\
	m_3^2 &=& (\mu_3^2 + 2 \lambda_2 v^2) - \frac{1}{2} \lambda_5 v^2, \nonumber \\
	m_5^2 &=& (\mu_3^2 + 2 \lambda_2 v^2) + \frac{1}{2} \lambda_5 v^2,
\end{eqnarray}
respectively.

When $\lambda_5$ is positive the real custodial-singlet neutral scalar with mass $m_1$ is the lightest of the $\mathbb{Z}_2$-odd states and is the dark matter candidate.  When $\lambda_5$ is negative the custodial-fiveplet states, comprising one neutral, one charged, and one doubly-charged scalar, are the lightest.
The question now is if we can have just one neutral state as the dark matter candidate. As shown in~\cite{Cirelli:2005uq} the mass degeneracy between charged and neutral states
can be lifted via quantum corrections,
which tend to make the charged components slightly heavier than the neutral ones. Therefore one would expect that in this phase of the GM model quantum corrections also
lift the degeneracy and that the neutral state would be the dark matter candidate. The model has a fully dark sector that only communicates with the visible sector via couplings to the SM Higgs and to the gauge bosons in quartic couplings. The visible sector is indistinguishable from the SM from the point of view of the (tree-level) Higgs couplings to fermions and gauge bosons, though the loop-induced coupling to photon pairs will be modified due to contributions from loops of the $\mathbb{Z}_2$-odd states.

Because $\mu_3^2$ remains a free parameter that can be taken arbitrarily large, this phase of the model possesses a decoupling limit, in which all of the $\mathbb{Z}_2$-odd states can be taken heavy with masses of order $\sqrt{\mu_3^2}$.

We end this section writing the X and $V'$ vectors for this stationary point, as defined in Eqs.~(\ref{eq:Xvector}) and (\ref{eq:Vl}),
\begin{equation}
	X_{DM}=\left(
	\begin{array}{c}
	m_h^2/(16 \lambda_1) \\
	0 \\
	0 \\
	0 \\
	0 \\
	0 \\
	\end{array}
	\right), \qquad 
	V'_{DM}=A + B X_{DM}=\left(
	\begin{array}{c}
	0 \\
	m_1^2+\bar{k}^2 \\
	\bar{k}^2 \\
	\bar{k}^2 \\
	\bar{k}^2 \\
	0 \\
	\end{array}
	\right) ,
\end{equation}
with $\bar{k}^2=-\frac{\lambda_5 m_h^2}{16 \lambda_1}.$

\section{Models where the custodial symmetry is broken after SSB}
\label{sec:noncust}

\subsection{The real non-custodial vacuum}

Let us now consider the scenario where the solution of the minimization equations has $\langle\varphi_8\rangle \neq \sqrt{2} \langle\varphi_5\rangle$
(but still no charged VEVs) which means that the custodial symmetry is broken. The absence of the dimension 3 soft breaking
terms will then lead to two massless scalars (the result of spontaneous breaking of continuous
global symmetries of the model). Let the VEV choice be
\begin{equation}
\langle \varphi_1\rangle = v_1\;,\;\langle \varphi_5\rangle = v_5\;,\;\langle \varphi_6\rangle = 0\;,\;
\langle \varphi_8\rangle = v_8\;,\; \langle \varphi_{10}\rangle=0 \;,\;
\label{eq:vevNONCUS}
\end{equation}
which leads to the minimization conditions
\begin{equation}
	\begin{aligned}
		\mu_2^2&=-4 \lambda_1 v_1^2-2 \lambda_2
		\left(v_5^2+v_8^2\right)+\frac{2 \lambda_3 v_8^2 \left(2
			\sqrt{2} v_5+v_8\right) \left(v_5 v_8^2-2
			v_5^3\right)}{v_1^2 \left(\sqrt{2} v_5^2+v_5
			v_8-\sqrt{2} v_8^2\right)}\, , \\
		\mu_3^2&=-2 \lambda_2 v_1^2-4 v_5^2 (\lambda_3+\lambda_4)-4
		v_8^2 (\lambda_3+\lambda_4)+\frac{4 \lambda_3 v_8^3}{\sqrt{2} v_5+2 v_8}\, ,\\
		\lambda_5&=\frac{4 \lambda_3 v_5 v_8 \left(v_8^2-2
			v_5^2\right)}{v_1^2 \left(\sqrt{2} v_5^2+v_5
			v_8-\sqrt{2} v_8^2\right)}\, .
	\end{aligned}
\label{eq:non-c}
\end{equation}

The vectors $X$ and $V^\prime$ are
\begin{equation}
X_{NC} = \left(
\begin{array}{c}
v_1^2/2 \\
1/2 \left(v_5^2+v_8^2\right) \\
\sqrt{2} v_5 v_8 \\
v_8^2/2 \\
0 \\
0 \\
\end{array}
\right)\, ,\quad
V^\prime_{NC}=\left(
\begin{array}{c}
0 \\
\frac{4 \lambda_3 v_8^3}{\sqrt{2} v_5+2 v_8} \\
-\frac{2 \lambda_3 v_5 v_8^2}{v_5+\sqrt{2} v_8}
\\
2 \lambda_3 v_8 \left(\frac{2 v_5^3-v_5
	v_8^2}{\sqrt{2} v_5^2+v_5 v_8-\sqrt{2}
	v_8^2}-v_8\right) \\
-\frac{2 \lambda_3 v_5 v_8^2}{v_5+\sqrt{2} v_8}\, ,
\\
\lambda_3 \left(-v_8^2\right) \\
\end{array}
\right)\, .
  \label{eq:xNCvlNC}
\end{equation}

\subsection{Charge breaking vacua}

We now consider the scenarios where charge breaking (CB) occurs spontaneously, and where therefore
the photon becomes massive after SSB. {\em A priori}, it would seem that there would be a large
number of possible CB vacua, but in fact -- as will be explained in appendix~\ref{app:vac} --
the minimization equations of the GM model curtail most of those possibilities. Indeed, there are
but five distinct vacua that lead to charge breaking,
\begin{enumerate}
	\item $v_5=v_8=0$ and $v_{10}=-\frac{\lambda_5}{\lambda_3} \frac{v_1^2}{8 v_6}$.
	\item $v_5=v_8=0$ and $v_{10}=- v_6$.
	\item $v_5=v_8=0$ and $v_{10}= v_6$.
	\item $v_1=v_5=v_8=0$ and $v_6=- v_{10}$.
	\item $v_1=v_5=v_8=0$ and $v_6= v_{10}$.
\end{enumerate}
To insist on this point, these are the {\em only} combinations of real VEVs originating
CB which are possible solutions of the extremum equations of the GM model.
The minimum conditions then lead to the following relations for vacuum number 1,
\begin{equation}
\begin{aligned}
	\mu_2^2&=-\frac{\lambda_2 \lambda_5^2 v_1^4}{32 \lambda_3^2 v_6^2}-\frac{v_1^2 \left(32 \lambda_1 \lambda_3+\lambda_5^2\right)}{8 \lambda_3}-2 \lambda_2\, ,\\
	\mu_3^2&=-\frac{\lambda_5^2 v_1^4 (\lambda_3+\lambda_4)}{16 \lambda_3^2 v_6^2}-2 \lambda_2 v_1^2-4
	v_6^2 (\lambda_3+\lambda_4) \, .
\end{aligned}
\end{equation}
For vacuum number 2 we have
\begin{equation}
	\begin{aligned}
	\mu_2^2&=-4 \lambda_1 v_1^2-v_6^2 (4 \lambda_2+\lambda_5)\, ,\\
	\mu_3^2&=-\frac{1}{2} v_1^2 (4 \lambda_2+\lambda_5)-4 v_6^2(\lambda_3+2 \lambda_4) \, .
	\end{aligned}
\end{equation}
For vacuum number 3 we obtain
\begin{equation}
\begin{aligned}
\mu_2^2&=-4 \lambda_1 v_1^2+v_6^2 (-4 \lambda_2+\lambda_5)\, ,\\
\mu_3^2&=\frac{1}{2} v_1^2 (-4 \lambda_2+\lambda_5)-4 v_6^2(\lambda_3+2 \lambda_4) \, .
\end{aligned}
\end{equation}
Finally for vacua 4 and 5 the condition is
\begin{equation}
	\mu_3^2=-4 v_6^2 (\lambda_3+2 \lambda_4) \, .
\end{equation}
It should be clear to the reader that the values of the VEVs shown here is almost certainly different
from vacuum to vacuum, though we are not labeling them differently.
Following the bilinear formalism, we can then write $X$ and $V'$ vectors for each CB vacuum:
\begin{itemize}
	\item Charge-breaking 1:
	\begin{equation}
	X_{CB1}=\left(
	\begin{array}{c}
	v_1^2/2 \\
	\bar{l}^2/2\\
	0 \\
	0 \\
	-v_1^2\lambda_5/(8\lambda_3)\\
	\bar{l}^2 \\
	\end{array}
	\right),  \qquad V'_{CB1}=A+ B X_{CB1}=\left(
	\begin{array}{c}
	0 \\
	0\\
	-v_1^2\lambda_5 /4 \\
	-v_1^2\lambda_5/2 - 2 \lambda_3\bar{l}^2\\
	0\\
	0 \\
	\end{array}
	\right) ,
	\end{equation}
	with $\bar{l}^2=v_6^2+\frac{v_1^4 \lambda_5^2}{64v_6^2 \lambda_3^2}$.
	
	\item Charge-breaking 2:
	\begin{equation}
	X_{CB2}=\left(
	\begin{array}{c}
	v_1^2/2 \\
	v_6^2 \\
	0 \\
	0 \\
	-v_6^2 \\
	2 v_6^2 \\
	\end{array}
	\right), \qquad V'_{CB2}=A + B X_{CB2}=\left(
	\begin{array}{c}
	0 \\
	4v_6^2\lambda_3-v_1^2\lambda_5/2\\
	2v_6^2\lambda_3-v_1^2\lambda_5/2 \\
	- 4v_6^2\lambda_3-v_1^2\lambda_5/2\\
	4v_6^2\lambda_3-v_1^2\lambda_5/2\\
	0 \\
	\end{array}
	\right) .
	\end{equation}
	
\item Charge-breaking 3:
\begin{equation}
X_{CB3}=\left(
\begin{array}{c}
v_1^2/2 \\
v_6^2 \\
0 \\
0 \\
v_6^2 \\
2 v_6^2 \\
\end{array}
\right), \qquad V'_{CB3}=A + B X_{CB3}=\left(
\begin{array}{c}
0\\
4v_6^2\lambda_3+v_1^2\lambda_5/2\\
-2v_6^2\lambda_3-v_1^2\lambda_5/2 \\
- 4v_6^2\lambda_3-v_1^2\lambda_5/2\\
-4v_6^2\lambda_3-v_1^2\lambda_5/2\\
0 \\
\end{array}
\right) .
\end{equation}

\item Charge-breaking 4:
\begin{equation}
X_{CB4}=\left(
\begin{array}{c}
0 \\
v_6^2 \\
0 \\
0 \\
-v_6^2 \\
2 v_6^2 \\
\end{array}
\right), \qquad V'_{CB4}=A + B X_{CB4}=\left(
\begin{array}{c}
\mu_2^2+v_6^2 (4 \lambda_2+\lambda_5) \\
4v_6^2\lambda_3 \\
2v_6^2\lambda_3 \\
-4v_6^2\lambda_3 \\
4v_6^2\lambda_3 \\
0 \\
\end{array}\right).
\end{equation}

\item Charge-breaking 5:
\begin{equation}
X_{CB5}=\left(
\begin{array}{c}
0 \\
v_6^2 \\
0 \\
0 \\
v_6^2 \\
2 v_6^2 \\
\end{array}
\right), \qquad V'_{CB5}=A + B X_{CB5}=\left(
\begin{array}{c}
\mu_2^2+v_6^2 (4 \lambda_2-\lambda_5) \\
4v_6^2\lambda_3 \\
-2v_6^2\lambda_3 \\
-4v_6^2\lambda_3 \\
-4v_6^2\lambda_3 \\
0 \\
\end{array}\right).
\end{equation}
	
\end{itemize}

\subsection{Wrong-Electroweak vacuum}

There is a set of unphysical 
solutions of the minimization equations that do not break the electroweak symmetry correctly, 
leading to a massless $Z$ boson -- a curious situation, only made possible by the possibility of
having VEVs for the triplet fields with zero hypercharge.
The two possible minima is this category are
\begin{itemize}
	\item \textbf{W1}: $	\langle \varphi_1\rangle = 0\;,\;\langle \varphi_5\rangle = 0\;,\;\langle \varphi_6\rangle = \frac{1}{2} \sqrt{-\frac{\mu_3^2}{\lambda_3+\lambda_4}}\;,\;
	\langle \varphi_8\rangle = 0\;,\; \langle \varphi_{10}\rangle=0 \;,\;
	\label{eq:vevWE1}$

	\item \textbf{W2}: $	\langle \varphi_1\rangle = 0\;,\;\langle \varphi_5\rangle = v_5\;,\;\langle \varphi_6\rangle = v_6\;,\;
	\langle \varphi_8\rangle = 0\;,\; \langle \varphi_{10}\rangle=0 \;,\;
	\label{eq:vevWE3}$
	
	with the further condition
	\begin{equation}
	\mu_3^2=-4 (\lambda_3+\lambda_4)
	\left(v_5^2+v_6^2\right) \, .
	\end{equation}
\end{itemize}
Notice, in fact, how the above stationary points correspond to VEVs for the fields $\xi^0$
and $\xi^+$ from \eqref{eq:PX}, which carry no hypercharge, allowing the $Z$ boson to remain
massless~\footnote{It may seem peculiar to have a VEV for $\xi^+$ and still have a massless photon
in these vacua, but in fact the definition of electric charge is different in this situation.}. Indeed,
the symmetry group left invariant after SSB for these vacua is $U(1)\times U(1)$ -- one of these $U(1)$ is
the unbroken hypercharge group, the other is ``extracted" from $SU(2)$.
	The $X$ and $V'$ vectors in this scenario are
	\begin{itemize}
		
		\item W1:
		\begin{equation}
		X_{W1}=\left(
		\begin{array}{c}
		0 \\
		-\frac{\mu_3^2}{8 (\lambda_3+\lambda_4)} \\
		0 \\
		0 \\
		0 \\
		-\frac{\mu_3^2}{4 (\lambda_3+\lambda_4)} \\
		\end{array}
		\right), \qquad V'_{W1}=A + B X_{W1}=\left(
		\begin{array}{c}
		\mu_2^2-\frac{\lambda_2 \mu_3^2}{2 (\lambda_3+\lambda_4)} \\
		0 \\
		0 \\
		\frac{\lambda_3 \mu_3^2}{2 (\lambda_3+\lambda_4)} \\
		0 \\
		0 \\
		\end{array}
		\right) \, .
		\end{equation}

		\item W2:
		\begin{equation}
		X_{W2}=\left(
		\begin{array}{c}
		0 \\
		\frac{1}{2} \left(v_5^2+v_6^2\right) \\
		0 \\
		0 \\
		0 \\
		v_6^2 \\
		\end{array}
		\right), \qquad V'_{W2}=A + B X_{W2}=\left(
		\begin{array}{c}
		\mu_2^2+2 \lambda_2 \left(v_5^2+v_6^2\right) \\
		0 \\
		0 \\
		-2 \lambda_3 \left(v_5^2+v_6^2\right) \\
		0 \\
		0 \\
		\end{array}
		\right) \, .
		\end{equation}
\end{itemize}

\subsection{Tantamount vacua}
There is a set of vacua that produces three massive gauge bosons and thus is tantamount to the SM 
electroweak breaking vacuum. The difference lies in the different gauge bosons' mass structure.
In the SM breaking we obtain $m_W=gv/2$ and $m_Z=\sqrt{g^2 + g'^2}v/2$ while in this case the neutral component of the $SU(2)_L$ gauge bosons remains massless (``the photon'') and the ``$Z$ boson'' is just the (massive) hypercharge gauge boson.
These vacua would therefore
yield an unrealistic value for the $\rho$ parameter, as well as unrealistic couplings of the $Z$ boson to fermions, and are thus in disagreement with current
 experimental data.

The vacua that produce these results are
\begin{itemize}
	\item \textbf{T1}: $	\langle \varphi_1\rangle = 0\;,\;\langle \varphi_5\rangle = v_5\;,\;\langle \varphi_6\rangle = 0\;,\;
	\langle \varphi_8\rangle = 0\;,\; \langle \varphi_{10}\rangle=v_{10} \;,\;
	\label{eq:vevTANT1}$
	with the condition
	\begin{equation}
		\mu_3^2=-4 (\lambda_3+\lambda_4)\left(v_{10}^2+v_5^2\right) \, .
	\end{equation}
In this vacuum the gauge boson masses are $m_W = g \sqrt{v_{10}^2 + v_5^2}$, $m_Z = g^{\prime} v_{10}$.
	
	\item \textbf{T2}: $	\langle \varphi_1\rangle = 0\;,\;\langle \varphi_5\rangle = 0\;,\;\langle \varphi_6\rangle = 0\;,\;
	\langle \varphi_8\rangle = 0\;,\; \langle \varphi_{10}\rangle=v_{10} \;,\;
	\label{eq:vevTANT2}$
		with the condition
	\begin{equation}
		\mu_3^2=-4 (\lambda_3+\lambda_4)v_{10}^2 \, .
	\end{equation}
In this vacuum the gauge boson masses are $m_W = g v_{10}$, $m_Z = g^{\prime} v_{10}$.

\end{itemize}

	The corresponding $X$ and $V'$ vectors are
	\begin{itemize}
		\item T1:
		\begin{equation}
			X_{T1}=\left(
			\begin{array}{c}
			0 \\
			-\frac{\mu_3^2}{8 (\lambda_3+\lambda_4)} \\
			0 \\
			0 \\
			0 \\
			-\frac{\mu_3^2}{4 (\lambda_3+\lambda_4)}-v_5^2 \\
			\end{array}
			\right),\qquad V'_{T1}=A + B X_{T1}=\left(
			\begin{array}{c}
			\mu_2^2-\frac{\lambda_2 \mu_3^2}{2
				(\lambda_3+\lambda_4)} \\
			0 \\
			0 \\
			\frac{\lambda_3 \mu_3^2}{2 (\lambda_3+\lambda_4)}+2 \lambda_3 v_5^2 \\
			0 \\
			0 \\
			\end{array}
			\right)\, .
		\end{equation}
		\item T2:
				\begin{equation}
		X_{T2}=\left(
		\begin{array}{c}
		0 \\
		-\frac{\mu_3^2}{8 (\lambda_3+\lambda_4)} \\
		0 \\
		0 \\
		0 \\
		-\frac{\mu_3^2}{4 (\lambda_3+\lambda_4)} \\
		\end{array}
		\right),\qquad V'_{T2}=A + B X_{T2}=\left(
		\begin{array}{c}
		\mu_2^2-\frac{\lambda_2 \mu_3^2}{2
			(\lambda_3+\lambda_4)} \\
		0 \\
		0 \\
		\frac{\lambda_3 \mu_3^2}{2 (\lambda_3+\lambda_4)} \\
		0 \\
		0 \\
		\end{array}
		\right)\, .
		\end{equation}

\end{itemize}

\section{Stability of the viable models}
\label{sec:viable}
\subsection{Stability of dark matter minima}

Let us now consider the coexistence of a dark matter {\em minimum} with any of the other stationary points
(minima or not) listed above. The starting point of this analysis is that the
minimization equations of the potential admit solutions of two different types
simultaneously. We force the dark matter solution to be a minimum while the stationary point to be compared has no constraints on the second derivatives.
The bilinear formalism provides an easy method to compute and
compare the values of the potential at each of those two stationary points.

Let us start by comparing the DM vacuum with all the CB ones. We will analyse in detail two of the cases,
the arguments will be the same for all the others.

\begin{itemize}
	\item CB1 vs. DM:
	\begin{itemize}
		\item for $\lambda_5>0$
		\begin{equation}
		V_{CB1}-V_{DM}=\frac{1}{2}(v_6^2+v_{10}^2)m_{H^+_1}^2+\frac{1}{4}v_1^2 \lambda_5(v_6-v_{10})^2>0\, ,
		\end{equation}
		\item for $\lambda_5<0$
		\begin{equation}
		V_{CB1}-V_{DM}=\frac{1}{2}(v_6^2+v_{10}^2)m_{H^+_2}^2-\frac{1}{4}v_1^2 \lambda_5(v_6+v_{10})^2>0\, ,
		\end{equation}
	\end{itemize}
	where all explicit VEVs are from the CB1 stationary point, and all masses are from the DM minimum. 
Given our initial hypothesis of the DM phase being a minimum, the squared scalar mass $m_{H^+_1}^2$
is necessarily positive; likewise, we see that all VEV combinations figuring in the above expressions
are perforce positive -- and the signs affecting the $\lambda_5$ coupling render all terms where it
appears also positive.
Thus, the difference between the potential at the different SPs, $V_{CB1}-V_{DM}$ is always positive --
this means that, if there is a DM minimum, any CB1 extremum will necessarily lie {\em above} it.
	
	\item CB2 vs. DM:
	\begin{equation}
	V_\text{CB2}-V_\text{DM}=\frac{m_{H^+_1}^2 v_6^2}{2}\, ,
	\end{equation}
	where all explicit VEVs are from CB2 and all masses are from the DM phase. This difference is clearly always positive when the DM phase is a minimum.
	
	\item CB3 vs. DM:
	\begin{equation}
	V_\text{CB3}-V_\text{DM}=\frac{m_{H^+_2}^2 v_6^2}{2}\, ,
	\end{equation}
	where again all explicit VEVs are from CB3 and all masses are from the DM phase. And this difference is again always positive.
	
	\item CB4 and CB5 vs. DM:
	\begin{equation}
	V_\text{CB4/5}-V_\text{DM}=\frac{m_h^4+32 \lambda_1 \mu_3^2 v_6^2}{64
		\lambda_1}\, ,
	\end{equation}
	where all explicit VEVs are from CB4 or CB5 and all masses are from the DM phase. This difference can
be either positive or negative -- notice how the quadratic coupling $\mu_3^2$ can be negative, and
therefore the sign of the potential difference between the two stationary points is not determined. In fact, a numerical
study of this case found regions of parameter space for which $V_\text{CB4/5}-V_\text{DM} > 0$, but also 
other regions where $V_\text{CB4/5}-V_\text{DM} < 0$.
	
	\item Custodial vacuum vs. DM:
	\begin{equation}
		V_\text{C}-V_\text{DM}=\frac{3 m_2^2 v_5^2}{4}\, ,
\label{eq:CDM}
	\end{equation}
	where all explicit VEVs are from the custodial vacuum and all masses are from the DM phase. This difference is thus always positive.
	
	\item Non-custodial vacuum vs. DM:
	\begin{equation}
		V_{NC}-V_{DM}=\frac{1}{12} \left[m_1^2 \left(\sqrt{2} v_5-v_8\right)^2+m_2^2
		\left(v_5+\sqrt{2} v_8\right)^2\right]\, ,
	\end{equation}
	where all explicit VEVs are from the non-custodial vacuum and the masses are from the DM minimum. This difference is thus always positive.
\end{itemize}
\begin{itemize}	
	
	\item W1, W2, T1, T2 vs. DM:
	\begin{equation}
		V_{W1, W2, T1, T2}-V_{DM}=\frac{1}{16} \left(\frac{\mu_2^4}{\lambda_1}-\frac{\mu_3^4}{\lambda_3+\lambda_4}\right)\, ,
\label{eq:WDM}
	\end{equation}
	The difference in values of the potential can be positive or negative, depending on the magnitudes of both quadratic coefficients\footnote{Notice that, due to the bounded from below conditions
of \eqref{eq:bfb}, the combinations of $\lambda$ coefficients in \eqref{eq:WDM} are
necessarily positive.}.
\end{itemize}

Therefore, taking into account all the conditions above, in order to force the DM vacuum to be an absolute minimum at tree-level we just need to impose the two conditions:
\begin{eqnarray}
	 \frac{8 \lambda_1 \mu_3^4}{\lambda_3 + 2 \lambda_4} & \leq & m_h^4  \label{eq:DMs1} \, , \\
	 \frac{\mu_2^4}{\lambda_1}-\frac{\mu_3^4}{\lambda_3+\lambda_4} & \geq &0 \, ,
\end{eqnarray}
where, remember, in the DM minimum $h$ is supposed to represent the SM-like Higgs boson and as such $m_h = 125$ GeV. Using \eqref{eq:vevdm} and~\eqref{eq:mhdm}, we can cast these two equations in a different 
form, as they become lower bounds on combinations of $\lambda_3$ and $\lambda_4$:
\begin{equation}
\mbox{min}\left(\frac{1}{2}\lambda_3 + \lambda_4 \,,\, \lambda_3 + \lambda_4 \right)\;\geq \;
\frac{\mu_3^4}{2 v^2 m^2_h}\,.
\label{eq:DMs2}
\end{equation}
Notice that the bounded from below conditions of \eqref{eq:bfb} force these two combinations
of  $\lambda_3$ and $\lambda_4$ to be positive, this constraint is therefore more restrictive\footnote{
Though this constraint presupposes the existence of some Wrong-electroweak or Tantamount stationary point coexisting with 
the DM minimum.}.

\subsection{Stability of the custodial vacuum}

As in the previous section for the DM phase, our starting point now will be to assume
the existence of a custodial {\em minimum}, coexisting with other types of stationary points. 
Let us go through the several possibilities:
\begin{itemize}
	\item DM vs. Custodial:
	\begin{equation}
		V_{DM}-V_{C}=\frac{3 v_1^2 v_5^2 (4 \lambda_1 \mu_3^2-2\lambda_2 \mu_2^2+\lambda_5\mu_2^2)^2}{4  \lambda_1 m_h^2 m_H^2} \, ,
\label{eq:DMC}
	\end{equation}
where please notice that the masses figuring in this expression are now those obtained in the
custodial vacuum (see \eqref{eq:mHh_C},~\eqref{eq:m3_C} and~\eqref{eq:m5_C}).
This difference in values of the potential at the two stationary points is clearly positive, which
indicates that, whenever there is a custodial minimum, any existing DM stationary point will necessarily lie {\em above} it. Notice, though, that we had already concluded, in \eqref{eq:CDM}, that a DM minimum implied that any custodial stationary point needed to lie above it. The two conclusions are not 
contradictory, rather they imply that if a DM (custodial) minimum exists, any coexisting 
custodial (DM) stationary point lies above it, and is necessarily {\em a saddle point}. In fact, comparing
\eqref{eq:CDM} and~\eqref{eq:DMC}, we see that when the minimum corresponds to the DM phase,
then $V_{DM}-V_{C} <0$ which, from the equation above, implies that {\em one} of the squared scalar masses
of the custodial stationary point, $m_h^2$ or $m_H^2$, must perforce be negative, while the other is positive -- thus we
conclude that not only the custodial stationary point lies above the DM minimum, it is also necessarily a 
{\em saddle point}. Analogously, if we have a custodial minimum then $V_{DM}-V_{C} > 0$ and from 
\eqref{eq:CDM}  we see that one must necessarily have one of the squared masses computed at the 
DM stationary point, $m_2^2$, be {\em negative} -- it is easy to then show that in that case the DM stationary point is also a saddle 
point, lying above the custodial minimum.
	
	\item Non-custodial vs. custodial:
	\begin{equation}
	V_{NC}-V_C= \frac{A^{NC/C}}{B^{NC/C}}
	\label{eq:NC1}
	\end{equation}
with
	\begin{eqnarray}
		A^{NC/C} &=&-\lambda_3 \left(2 v_5^2-v_8^2\right)
			\left\{ -v_1^4 \left(-32 v_5^6 v_8^2-4
			\sqrt{2} v_5^5 v_8^3+50 v_5^4 v_8^4-11
			\sqrt{2} v_5^3 v_8^5-26 v_5^2 v_8^6\right. \right. 
			\nonumber \\&& \left. 
			+ 4\sqrt{2} v_5^7 v_8+8 v_5^8+14 \sqrt{2}
			v_5 v_8^7-4 v_8^8\right)  \nonumber \\
			&& 
			\times \left\{v_5^2 \left[\lambda_2^2-4 \lambda_1 (\lambda_3+\lambda_4)\right]-2 v_8^2 \left[\lambda_2^2-2\lambda_1 (\lambda_3+2 \lambda_4)\right]\right\} \nonumber \\
			&& \left.
			+4	\lambda_2 \lambda_3 v_1^2 v_5 v_8^2 \left(2 v_5^2-v_8^2\right) \right.\nonumber \\&&
			\times \left(-12
			v_5^5 v_8^2-8 \sqrt{2} v_5^4 v_8^3+17
			v_5^3 v_8^4+3 \sqrt{2} v_5^2 v_8^5+4
			\sqrt{2} v_5^6 v_8+4 v_5^7-10 v_5
			v_8^6+2 \sqrt{2} v_8^7\right) \nonumber \\&&
			 +2 \lambda_3
			v_5^2 v_8^2 \left(2 v_5^2-v_8^2\right)
			\left[24 v_5^8 (\lambda_3+\lambda_4)-16 v_5^6 v_8^2 (5 \lambda_3+6
			\lambda_4)-12 \sqrt{2} v_5^5 v_8^3
			(\lambda_3+\lambda_4) \right.
\nonumber \\&&			
			+6 v_5^4 v_8^4 (17 \lambda_3+25\lambda_4)-\sqrt{2} v_5^3 v_8^5 (17 \lambda_3+33 \lambda_4)
\nonumber \\&&			-6 v_5^2 v_8^6 (7 \lambda_3+13\lambda_4)+12 \sqrt{2}	v_5^7 v_8 (\lambda_3+\lambda_4)
\nonumber \\
&& \left. \left.
+ 6 \sqrt{2} v_5 v_8^7 (3 \lambda_3+7\lambda_4)-4 v_8^8 (\lambda_3+3\lambda_4)\right] \right\} \, ,
	\label{eq:VNC_VC1}
\end{eqnarray}
\begin{eqnarray}
B^{NC/C} &=& 2 \sqrt{2} \left(v_5+\sqrt{2} v_8\right)^2 \left(\sqrt{2} v_5^2+v_5 v_8-\sqrt{2} v_8^2\right) \nonumber \\
&& \times \left\{v_1^4	\left(-3 v_5^2 v_8^2+2 \sqrt{2} v_5^3 v_8+2 v_5^4-2 \sqrt{2} v_5 v_8^3+2 v_8^4\right)
\left[3 \lambda_2^2-4 \lambda_1 (\lambda_3+3 \lambda_4)\right] \right. \nonumber \\
&& \left.
+12 \lambda_2 \lambda_3 v_1^2 v_5 v_8 \left(2	v_5^2-v_8^2\right) \left(\sqrt{2}v_5^2+v_5 v8-\sqrt{2} v_8^2\right)  +12	\lambda_3^2 v_5^2 v_8^2	\left(v_8^2-2 v_5^2\right)^2\right\} \, .\qquad
	\label{eq:VNC_VC2}
	\end{eqnarray}
The VEVs are from the NC vacuum. The difference can be positive or negative.

\item CB1 vs. Custodial:
\begin{equation}
	V_{CB1}-V_C= \frac{A^{CB1/C}}{B^{CB1/C}},
		\label{eq:CBC1}
	\end{equation}
with
	\begin{eqnarray}
		       && A^{CB1/C} = \mu_3^4 \left[1024 \lambda_1^2
			\lambda_3^2+128 \lambda_1 \lambda_3
			\lambda_5 (\lambda_5-3 \lambda_2)+3
			\lambda_5^2 (\lambda_5-2 \lambda_2)^2\right] \nonumber\\&&
			+8 \mu_2^2 \mu_3^2 \left\{32
			\lambda_1 \lambda_3 \left[3 \lambda_5
			(\lambda_3+\lambda_4)-4 \lambda_2
			\lambda_3\right]+3 \lambda_5 (2 \lambda_2-\lambda_5) \left[4 \lambda_2 \lambda_3-\lambda_5 (\lambda_3+\lambda_4)\right] \right\}\nonumber \\ &&
			+16 \mu_2^4 \left[16 \lambda_2^2
			\lambda_3^2-24 \lambda_2 \lambda_3
			\lambda_5 (\lambda_3+\lambda_4)+\lambda_5^2 (\lambda_3+\lambda_4) (7 \lambda_3+3 \lambda_4)\right]\, , \qquad  \label{eq:CBCA1}
	\end{eqnarray}
		\begin{eqnarray}			%
			&& B^{CB1/C} =16 \left[3 (\lambda_5-2 \lambda_2)^2-16
			\lambda_1 (\lambda_3+3 \lambda_4)\right] \left[8 \lambda_2^2 \lambda_3-(\lambda_3+\lambda_4) \left(32
			\lambda_1 \lambda_3+\lambda_5^2\right)\right] \, . \qquad  \label{eq:CBCB1}
	\end{eqnarray}
	
	 Numerically, we found that a coexisting CB1 vacuum can have an energy above or below the Custodial vacuum.
	
	\item CB2 vs. Custodial:
	\begin{equation}
		V_{CB2}-V_{C}=\frac{1}{8} \left[\frac{9 m_3^2 \left(v_{C,1}^2 v_{CB,6}^2+2 v_{CB,1}^2 v_{C,5}^2\right)}{v_{C,1}^2+8 v_{C,5}^2}+m_5^2 v_{CB,6}^2\right]\, ,
	\end{equation}
	where VEVs are marked as either being charge-breaking or custodial and all masses are from the custodial vacuum. The difference is thus always positive.
	
	\item CB3 vs. Custodial:
	\begin{equation}
		V_{CB3}-V_{C}=\frac{1}{8} \left[\frac{m_3^2 \left(v_{C,1}^2 v_{CB,6}^2+6 v_{CB,1}^2
		v_{C,5}^2\right)}{v_{C,1}^2+8 v_{C,5}^2}+m_5^2 v_{CB,6}^2\right]\, ,
	\end{equation}
	where VEVs are marked as either being charge-breaking or custodial and all masses are from the custodial vacuum. The difference is always positive.
	
	\item CB4/5 vs. Custodial:
	\begin{equation}
		V_{CB4/5}-V_{C}=-\frac{v_1^2 (-2 \lambda_2 \mu_3^2+2
			\lambda_3 \mu_2^2+4 \lambda_4
			\mu_2^2+\lambda_5 \mu_3^2)+2
			\lambda_3 \mu_3^2 v_5^2}{8
			(\lambda_3+2 \lambda_4)}\, ,
	\end{equation}
	where VEVs are from the custodial vacuum. The difference can be positive or negative.
	
\end{itemize}
	
\begin{itemize}
	\item W1, W2, T1 and T2 vs. Custodial. For all these scenarios the difference in height in the potential is given by the expression
	\begin{equation}
		V_{W1, W2, T1, T2}-V_{C}=\frac{v^2 \left(2 \mu_3^2 v_5^2 \left(m_5^2-2 \lambda_2 v_1^2\right)+\mu_2^2 v_1^2
			\left(m_5^2+8 \lambda_4
			v_5^2\right)\right)-m_3^2 \left(3 \mu_2^2
			v_1^4+2 \mu_3^2 v_1^2
			v_5^2\right)}{12 m_3^2 v_1^2-4
			v^2 \left(m_5^2+8
			\lambda_4 v_5^2\right)}\, ,
	\end{equation}
	where VEVs and masses are from the Custodial vacuum. The difference can be positive or negative.	
\end{itemize}

Therefore, we can finally state the conditions to have the custodial extremum as the absolute minimum. 
Besides having all masses positive to guarantee that we are
indeed in a minimum, we need to force the parameters of the potential to comply with the following conditions
\begin{eqnarray}
&&	  V_{NC}-V_C  \geq  0, \label{eq:C1}\\
&&	  V_{CB1}-V_C  \geq 0, \label{eq:C2} \\
&&	  	  -\frac{v_1^2 (-2 \lambda_2 \mu_3^2+2
			\lambda_3 \mu_2^2+4 \lambda_4
			\mu_2^2+\lambda_5 \mu_3^2)+2
			\lambda_3 \mu_3^2 v_5^2}{8
			(\lambda_3+2 \lambda_4)}  \geq 0, \label{eq:C3}\\
&&		  	  \frac{v^2 \left(2 \mu_3^2 v_5^2 \left(m_5^2-2 \lambda_2 v_1^2\right)+\mu_2^2 v_1^2
			\left(m_5^2+8 \lambda_4
			v_5^2\right)\right)-m_3^2 \left(3 \mu_2^2
			v_1^4+2 \mu_3^2 v_1^2
			v_5^2\right)}{12 m_3^2 v_1^2-4
			v^2 \left(m_5^2+8
			\lambda_4 v_5^2\right)}  \geq 0. \label{eq:C4}
\end{eqnarray}
The first equation, \eqref{eq:C1}, is written using \eqref{eq:NC1}, \eqref{eq:VNC_VC1} and \eqref{eq:VNC_VC2}.
They are written
as a function of the non-custodial VEVs that obey Eqs. (\ref{eq:non-c}). The remaining three equations  
\eqref{eq:C2},  \eqref{eq:C3},  \eqref{eq:C4}
are written as a function of the parameters of the potential, the masses in the custodial phase and the 
custodial VEVs. In particular, \eqref{eq:C2}
needs  \eqref{eq:CBC1}, \eqref{eq:CBCA1} and \eqref{eq:CBCB1}.

\section{Conclusions}
\label{sec:Conc}

We have analysed the vacuum structure of a simple version of the GM model where the scalar potential is not only invariant
under $SU(2)_L \times SU(2)_R$ but is also invariant under  the $\mathbb{Z}_2$ symmetry of the triplet fields, $\Xi \to -\Xi$.
We have then used the invariance freedom to reduce the number of possible VEVs to 7. In this first attempt to understand the
vacuum structure of the model we have further reduced the number to 5 by considering only the real component of the VEVs.
In that sense the conditions obtained for the potential to be in an absolute minimum are necessary conditions.

We have classified the phases of the model in viable phases, the Custodial and DM phases, and the non-viable. The non-viable
are the non-custodial vacuum where 5 Goldstone bosons are generated, the charge-breaking vacua (CB1,2 with 6 Goldstone bosons
and CB3,4,5 with 5 Goldstone bosons), and the wrong-electroweak and tantamount vacua with 4 Goldstone bosons.
We have then derived a set of analytical formulae that allows to compare the difference in the depths of two stationary points from two distinct phases. If we
want to be certain that we have an absolute minimum in the DM phase, we just need that all masses be positive in that phase and to impose the conditions
given in equation  (\ref{eq:DMs2}). These are the two conditions for which we have checked numerically that the difference
in depths could be either positive or negative. Hence, we need to force the DM minimum to be below the other phases' stationary point.
If in turn we want the custodial vacuum to be a minimum we impose positivity of the masses in that phase together with the conditions expressed by
equations (\ref{eq:C1}),  (\ref{eq:C2}), (\ref{eq:C3}) and (\ref{eq:C4}). Note that there are many cases where the minimum is always below the stationary point
of the other phase at tree-level. This is for instance the case where the DM minimum is compared with the Custodial stationary point or when the Custodial minimum is compared with
the DM stationary point.

The Georgi-Machacek model has therefore an elaborate vacuum structure -- vacua with Dark Matter
candidates; vacua which preserve or not custodial symmetry; and patterns of electroweak breaking with 
remaining symmetry group $U(1)\times U(1)$, with massless photon and $Z$, or a wrong value predicted for the
Weinberg angle. Crucially, though, this foray into the many possible vacua of the model revealed that neither
of the vacua of interest -- the Dark Matter minimum or the minimum which preserves custodial symmetry -- 
are automatically stable, as occurs for instance in the 
2HDM~\cite{Ferreira:2004yd,Barroso:2005sm,Ivanov:2006yq,Ivanov:2007de}, a model for which it was proven that
minima which break different symmetries cannot coexist. For the GM model, though, we have shown that, though 
the Dark Matter minimum cannot coexist with deeper Charge Breaking, Custodial or Non-Custodial vacua, it can
nonetheless have deeper stationary points with incorrect electroweak symmetry breaking (yielding massless $Z$'s or
the wrong value of $\theta_W$). Further, the Custodial minimum is not safe from eventual tunnelling 
to a deeper charge breaking vacuum of the type CB1, or indeed to a non-custodial vacuum (this last 
conclusion is known, see~\cite{Hartling:2014zca,Hartling:2014xma}). The situation will of course become 
even more complex when one starts analysing the GM model with soft breaking terms of the $\mathbb{Z}_2$
considered in this paper -- the more common form of the GM model, studied for instance 
in~\cite{Hartling:2014zca}. Such terms will obviously change the possible vacua and, being cubic in the
fields, will necessarily complicate the algebra. Many of the stranger vacua discussed here -- the 
Wrong-Electroweak and Tantamount ones -- may well no longer be possible when soft breaking terms appear in 
the potential. On the other hand, taking the 2HDM as an example, new types minima may become possible
with those soft breaking terms (such as for instance spontaneous CP breaking vacua), and coexistence of minima
impossible with exact symmetries may be possible when they are softly 
broken~\cite{Barroso:2007rr,Ivanov:2007de}.


\appendix
\section{Vacua}
\label{app:vac}

We have presented the simplest and most general vacuum structure in~\eqref{eq:vevs2}. We will now show the possible types of vacua in the $\mathbb{Z}_2$ symmetric
model. The most general vacuum configuration is written as
\begin{equation}
\Phi = \frac{1}{\sqrt{2}}\left( \begin{array}{cc}
v_1 & 0  \\
0 & v_1  \end{array} \right), \qquad \qquad
X =\frac{1}{\sqrt{2}}
\left(
\begin{array}{ccc}
v_8  & v_6 & 0 \\
-v_{10} & \sqrt{2} v_5 & v_{10} \\
0 & -v_6 & v_8
\end{array}
\right) \, ,
\label{eq:vevs_real}
\end{equation}
and leads to the following minimum conditions:
\begin{eqnarray}
	\frac{\partial V}{\partial \varphi_1}\Bigg|_0 &= &v_1 \left[\mu_2^2+4 \lambda_1 v_1^2 + 2 \lambda_2 (v_5^2+v_6^2+v_8^2+v_{10}^2)-\lambda_5 (v_{10} v_6 +\sqrt{2}  v_5)
	v_8 + \frac{v_8^2}{2}\right],  \\
	\frac{\partial V}{\partial \varphi_3}\Bigg|_0 &=&-\frac{ \lambda_5 }{\sqrt{2}}v_1 v_8 (v_6+v_{10}), \\
	 \frac{\partial V}{\partial \varphi_5}\Bigg|_0&=& v_5 \left[\mu_3^2+4 \lambda_4
	v_8^2+4( v_5^2+v_6^2+v_{10}^2)
	(\lambda_3+\lambda_4)-2 \sqrt{2}
	\lambda_3 \frac{v_6 v_8 v_{10}}{v_5} \right. \nonumber \\
	&& \left. +v_1^2 \left(2 \lambda_2
	-\frac{\lambda_5
		v_8}{\sqrt{2}v_5}\right) \right], \\
	\frac{\partial V}{\partial \varphi_6}\Bigg|_0&= & v_6 \left[ \mu_3^2+4 \lambda_4
	v_{10}^2 +4 (v_5^2+v_6^2)
	(\lambda_3+\lambda_4)+2
	v_8^2 (\lambda_3+2 \lambda_4)-2 \sqrt{2}
		\lambda_3 \frac{ v_5 v_8 v_{10}}{v_6} \right. \nonumber \\
		&& \left.
		+ v_1^2 \left(2 \lambda_2-\frac{\lambda_5
		v_{10}}{2 v_6}\right)\right], \\
	\frac{\partial V}{\partial \varphi_8}\Bigg|_0&=& v_8 \left\{ \mu_3^2+4 \lambda_4 v_5^2+2 (v_6^2+v_8^2+v_{10}^2) (\lambda_3+2
	\lambda_4)-2 \sqrt{2}
	\lambda_3 \frac{v_5 v_6 v_{10}}{v_8} \right. \nonumber \\
	&& \left.+ v_1^2 \left[2 \lambda_2 -\left(1+\sqrt{2}\frac{v_5}{v_8}\right) \frac{\lambda_5}{2} \right] \right\}, \\
	\frac{\partial V}{\partial \varphi_{10}}\Bigg|_0&=& v_{10} \left[ \mu_3^2+4 \lambda_4 v_6^2+4 (v_5^2+v_{10}^2) (\lambda_3+\lambda_4)+2 v_8^2
	(\lambda_3+2 \lambda_4)-2 \sqrt{2} \lambda_3
	\frac{v_5 v_6 v_8}{v_{10}}+ \right. \nonumber \\
	&& \left.
	v_1^2 \left(2 \lambda_2 -\frac{\lambda_5
		v_6}{2 v_{10}}\right)\right], \\
	\frac{\partial V}{\partial \varphi_{11}}\Bigg|_0&=&2 \lambda_3 \left(\sqrt{2}  v_5
	v_6 v_{10}- v_8v_{10}^2-v_6^2 v_8\right),
\end{eqnarray}
and the conditions for the remaining fields are identically zero.

The minimum conditions for $\varphi_3$ can be used as a starting point of three main subgroups with either $v_1=0$, $v_8=0$ or $v_6=-v_{10}$.
We then found step by step all configurations that led to stationary points of the potential.
All charge breaking minima have in common the following conditions
\begin{equation}
	v_5=v_8=0, \qquad v_6\neq 0, \text{ and } v_{10}\neq 0,
\end{equation}
that is, the neutral VEVs from the triplet have to vanish while the charged ones have to be non-zero. There are also Charge-Breaking minima
that obey all the above conditions plus $v_1=0$. In principle there could be other vacuum configurations that would break
electric charge but we found that they are never stationary points. By taking $v_5=v_8=0$ only three stationarity equations survive,
the ones related to $\varphi_1$, $\varphi_6$ and $\varphi_{10}$. This imposes conditions on three (arbitrary) Lagrangian parameters, and we choose
\begin{equation}
	\begin{aligned}
	\mu_2^2&=-4 \lambda_1 v_1^2-\frac{8 \lambda_3 v_{10}^2
		v_6^2}{v_1^2}-2 \lambda_2
	\left(v_{10}^2+v_6^2\right)\, ,\\
	\mu_3^2&=-2 \left(\lambda_2 v_1^2+2 (\lambda_3+\lambda_4)
	\left(v_{10}^2+v_6^2\right)\right) \, , \\
	\lambda_5&=-\frac{8 \lambda_3 v_{10} v_6}{v_1^2} \, .
	\end{aligned}
\end{equation}

The condition $v_5=v_8=0$ together with $v_6\neq 0 \text{ and } v_{10}\neq 0$ leads to the following mass matrix for the gauge bosons in the $W^1,W^2,W^3,B$ basis.
\begin{equation}
	\left(
	\begin{array}{cccc}
	\frac{1}{4} g^2 \left(v_1^2+4 v_{10}^2\right) & 0 & 0 & 0 \\
	0 & \frac{1}{4} g^2 \left(v_1^2+4
	\left(v_6^2+v_{10}^2\right)\right) & 0 & 0 \\
	0 & 0 & \frac{1}{4} g^2 \left(v_1^2+4 v_6^2\right) &
	-\frac{1}{4} v_1^2 g g' \\
	0 & 0 & -\frac{1}{4} v_1^2 g g' & \frac{1}{4} g'^2
	\left(v_1^2+4 v_{10}^2\right) \\
	\end{array}
	\right)
\end{equation}
Clearly there is no zero eigenstate and thus the photon becomes massive. There is another Charge-Breaking stationary point with the conditions $v_1= v_5=v_8=0$ together with $v_6\neq 0 \text{ and } v_{10}\neq 0$.
Again for this case we see that no zero mass eigenstate emerges.



\subsubsection*{Acknowledgments}
DA, PF and RS are supported by FCT, Contracts UIDB/00618/2020, UIDP/00618/2020, PTDC/FIS-PAR/31000/2017, CERN/FISPAR/0002/2017, CERN/FIS-PAR/0014/2019, and by the HARMONIA project, contract UMO-2015/18/M/ST2/0518.  HEL is supported by the Natural Sciences and Engineering Research Council of Canada (NSERC).  This work was also supported by the grant H2020-MSCA-RISE-2014 No.\ 645722 (NonMinimalHiggs).

\bibliography{GM}
\bibliographystyle{jhep}

\end{document}